\documentclass[prl,twocolumn,preprintnumbers,amsmath,amssymb]{revtex4}

\usepackage{graphicx}
\usepackage{bm}
\usepackage{color}

\begin{document}


\title{Revisiting the low temperature electron-phonon relaxation of noble metals}
\author{Shota Ono}
\email{shota\_o@gifu-u.ac.jp}
\affiliation{Department of Electrical, Electronic and Computer Engineering, Gifu University, Gifu 501-1193, Japan}

\begin{abstract}
The low temperature electron-phonon (e-ph) relaxation near the surface of noble metals, Cu and Ag, is studied by using the density-functional theory approach. The appearance of the surface phonon mode can give rise to a strong enhancement of the Eliashberg function at low frequency $\omega$. Assuming the Eliashberg function proportional to the square of $\omega$ in the low frequency limit, the e-ph relaxation time obtained from the surface calculations is shorter than that from the bulk calculation. The calculated e-ph relaxation time for the former is in agreement with a recent experiment for thin films. 
\end{abstract}


\maketitle

{\it Introduction.---}The electron-phonon (e-ph) interaction, one of the most important concepts in the many-body theory, accounts for the superconductivity and the electrical resistivity \cite{ziman,grimvall}. Eliashberg function is a key to understand the details of the e-ph interaction in a variety of metals. For example, the integration of the weighted Eliashberg function results in the e-ph coupling constant that determines the superconducting transition temperature. 

The e-ph coupling also plays an important role in nonequilibrium condition between electrons and phonons. The energy transfer dynamics is approximately described by the two-temperature model (TTM) for the electron temperature $T_{\rm e}$ and the phonon temperature $T_{\rm p}$ \cite{allen}. The temperature $(T)$ relaxation time $\tau$ in the TTM is different between the high and the low-$T$ limit. At high $T$, where the thermal energy is much larger than the Debye energy, $\tau$ is proportional to $T$. This model has been used in the field of ultrafast dynamics \cite{broson,demsar,kabanov}, while several models beyond the TTM have been proposed recently \cite{kabanov,waldecker,maldonado,ono2018}. At low $T$, it crucially depends on the low frequency ($\omega$) behavior of the Eliashberg function: When the behavior $\omega^p$ with an integer $p$ is assumed in the low $\omega$ limit, $\tau\propto T^{-p-1}$. For clean metals, $p=2$ \cite{wellstood}, while for dirty metals, the value of $p$ is scattered \cite{bergmann,echternach,sergeev,karvonen}. It also depends on the dimension and the boundary condition in a complicated manner \cite{qu,hekking,cojocaru}.

Notwithstanding the importance in the field of thermometry \cite{review}, the understanding of the low $T$ e-ph relaxation is still under debate even for clean metals. The magnitude of $\tau$ can be related to the energy transfer rate $Q = \Sigma \Omega (T_{\rm e}^{5} - T_{\rm ph}^{5})$, where $\Omega$ is the volume of the metal and $\Sigma$ is a parameter involving the e-ph coupling introduced in Ref.~\cite{wellstood}. The magnitude of $\Sigma$ evaluated within the deformation potential (DP) approximation has been found to be much smaller than that measured experimentally. This may be attributed to a lack of the umklapp scattering contributuion \cite{wellstood} and the non-spherical effect of the Fermi surface \cite{qu} in the DP model. However, a recent experiment has shown that the free-electron model can well describe the e-ph relaxation in Ag but Cu \cite{viisanen2018}. The density-functional theory (DFT) approach must enable us to calculate the Eliashberg function at low $\omega$ accurately and therefore to investigate $\tau$ at low $T$. However, to the best of our knowledge, such a DFT approach has yet to be applied to this issue.

At low $T$, the electron mean free path becomes large, implying that the electron scattering via the surface phonons would be important. In this paper, by performing DFT calculations, we investigate the low $T$ e-ph relaxation near the surface as well as bulk of noble metals, Cu and Ag. Assuming the Eliashberg function proportional to $\omega^2$, we show that the magnitude of $\tau$ for the surface model is about three times shorter than that for bulk and is in agreement with experiment \cite{viisanen2018}. We demonstrate that the electron-surface phonon interaction gives rise to an enhancement of the Eliashberg function at low $\omega$ and plays a key role to interpret the low $T$ e-ph dynamics.  

{\it Basic concepts.---}The energy transfer rate of the total electron energy $E_{\rm e}$ per unit cell is derived from the Boltzmann equation for the electron and phonon distribution functions under the assumption: (i) The effects of the diffusion and the external forces are neglected; (ii) The electron and phonon quasiequilibrium is established at any time. Using the Sommerfeld expansion $E_{\rm e}(T_{\rm e}) = E_{\rm e}(0) + \gamma T_{\rm e}^{2}/2$ with $\gamma = 2\pi^2 N_{\rm F}N_{\rm c}k_{\rm B}^2/3$, the time ($t$)-evolution of $T_{\rm e}$ is given by \cite{allen}
\begin{eqnarray}
C_{\rm e}\frac{dT_{\rm e}}{dt} &=& - \Gamma(T_{\rm e}) + \Gamma(T_{\rm p}),
\label{eq:rate_temp}
 \\
\Gamma(T) &=& 
 4\pi N_{\rm F}N_{\rm c}
 \int_{0}^{\omega_{\rm D}} d\omega (\hbar\omega)^2 
 \alpha^2 F(\omega) 
n_{\rm B}(\omega,T),
\label{eq:rate_gamma}
\end{eqnarray}
where $C_{\rm e}=\gamma T_{\rm e}$ is the specific heat of the electron, $N_{\rm F}$ is the electron density-of-states (DOS) per unit cell and per spin at the Fermi energy $\varepsilon_{\rm F}$, $N_{\rm c}$ is the number of unit cell, $n_{\rm B}(\omega,T)$ is the Bose-Einstein function at temperature $T$, and $\omega_{\rm D}$ is the Debye frequency. The Eliashberg function $\alpha^2 F(\omega)$ is given by
\begin{eqnarray}
\alpha^2 F(\omega)
 &=& \frac{1}{\hbar N_{\rm F}N_{\rm c}}
 \sum_{n,n',\bm{k},}\sum_{\gamma,\bm{q}} \vert g_{n,n'}^{\gamma}(\bm{k},\bm{q}) \vert^2
 \nonumber\\
 &\times&
 \delta (\varepsilon_{\rm F} - \varepsilon_{n\bm{k}})
   \delta (\varepsilon_{\rm F} - \varepsilon_{n'\bm{k}+\bm{q}})
    \delta (\omega - \omega_{\gamma\bm{q}}),
    \label{eq:coupling_e-ph}
\end{eqnarray}
where $\varepsilon_{n\bm{k}}$ is the single-particle electron energy with the wavevector $\bm{k}$ and the band index $n$, $\omega_{\gamma\bm{q}}$ is the phonon frequency for the wavevector $\bm{q}$ and the branch index $\gamma$, and $g_{n,n'}^{\gamma}(\bm{k},\bm{q})$ is the matrix elements for the e-ph interaction Hamiltonian. The $p$th moment of $\alpha^2 F(\omega)$ is defined as
\begin{eqnarray}
 \lambda\langle \omega^p \rangle &=&
 2 \int d\omega  \alpha^2 F(\omega) \omega^{p-1}
\end{eqnarray}
with an integer $p$. $\lambda$ and $\lambda\langle \omega^2 \rangle$ have been used as a measure of the strength of the e-ph coupling in metals.

\begin{table*}
\begin{center}
\caption{The values of $\Omega_{\rm cell}$ (\AA$^3$), $\lambda$, $\lambda\langle\omega^2 \rangle$ (meV$^2$), $N_{\rm F}$ (states/eV/spin/unit cell), $G$ ($10^{-4}$), $\Sigma_{\rm low}$ (GW/m$^3$/K$^5$), and $\tau_{\rm low}(T=0.1 {\rm {K}})$ ($\mu$s). }
{
\begin{tabular}{lccccccc}\hline
   & $\Omega_{\rm cell}$ \hspace{5mm} & $\lambda$\hspace{5mm} & $\lambda\langle \omega^2 \rangle$ \hspace{5mm} & $N_{\rm F}$ \hspace{5mm} & $G$ \hspace{5mm} & $\Sigma_{\rm low}$ \hspace{5mm} & $\tau_{\rm low}$ \\ \hline
 Cu bulk              \hspace{5mm} &   12.02 \hspace{5mm} & 0.13 \hspace{5mm} & 50.5 \hspace{5mm} & 0.15 \hspace{5mm} & 0.6 \hspace{5mm} & 0.28 \hspace{5mm} & 72.1 \\ \hline
 Cu (001) 5 layer \hspace{5mm} & 133.7 \hspace{5mm} & 0.13 \hspace{5mm} & 42.6 \hspace{5mm} & 0.78 \hspace{5mm} & 1.5 \hspace{5mm} & 0.31 \hspace{5mm} & 29.3 \\ \hline
 Cu (001) 7 layer \hspace{5mm} & 201.9 \hspace{5mm} & 0.14 \hspace{5mm} & 50.0 \hspace{5mm} & 1.10 \hspace{5mm} & 1.5 \hspace{5mm} & 0.29 \hspace{5mm} & 29.4 \\ \hline
 Cu (111) 7 layer \hspace{5mm} & 202.7 \hspace{5mm} & 0.13 \hspace{5mm} & 45.4 \hspace{5mm} & 1.12 \hspace{5mm} & 1.5 \hspace{5mm} & 0.30 \hspace{5mm} & 29.4 \\ \hline
 Ag bulk              \hspace{5mm} &   17.92 \hspace{5mm} & 0.14 \hspace{5mm} & 24.4 \hspace{5mm} & 0.14 \hspace{5mm} & 2.0 \hspace{5mm} & 0.55 \hspace{5mm} & 21.6 \\ \hline
 Ag (001) 5 layer \hspace{5mm} & 200.4 \hspace{5mm} & 0.15 \hspace{5mm} & 20.8 \hspace{5mm} & 0.68 \hspace{5mm} & 6.0 \hspace{5mm} & 0.73 \hspace{5mm} & 7.2  \\ \hline
 Ag (001) 7 layer \hspace{5mm} & 301.8 \hspace{5mm} & 0.15 \hspace{5mm} & 22.2 \hspace{5mm} & 0.97 \hspace{5mm} & 6.0 \hspace{5mm} & 0.69 \hspace{5mm} & 7.2 \\ \hline
 Ag (111) 7 layer \hspace{5mm} & 302.5 \hspace{5mm} & 0.14 \hspace{5mm} & 21.2 \hspace{5mm} & 0.97 \hspace{5mm} & 6.0 \hspace{5mm} & 0.69 \hspace{5mm} & 7.2\\ \hline
\end{tabular}
}
\label{table1}
\end{center}
\end{table*}

At high $T$, where $n_{\rm B}(\omega,T)\simeq k_{\rm B}T/(\hbar\omega)$ holds, $\Gamma (T)$ can be expressed by $\lambda\langle \omega^2 \rangle$. At low $T$, where the thermal energy $k_{\rm B}T$ is much smaller than the Debye energy $\hbar\omega_{\rm D}$, numerical integration of Eq.~(\ref{eq:rate_gamma}) must be performed by using an explicit expression of $\alpha^2 F(\omega)$. In a clean metal, it is expected to be $\alpha^2 F(\omega) \propto \omega^2$ at low $\omega$, equivalently $\Gamma (T)$ obeys a $T^5$ law. We thus define the e-ph coupling factor $\Sigma_{\rm low}$ as
\begin{eqnarray}
\Sigma_{\rm low} =  \frac{\Gamma_{\rm low} (T)}{N_{\rm c}{\Omega_{\rm cell}} T^{5}},
 \label{eq:gamma5}
\end{eqnarray}
where $\Omega_{\rm cell}$ is the volume of a unit cell.


From Eq.~(\ref{eq:rate_temp}), the time-evolution for $T_{\rm e}$ can be written as
\begin{eqnarray}
\frac{dT_{\rm e}}{dt} &=& - \frac{1}{\tau}(T_{\rm e}-T_{\rm p}).
\end{eqnarray}
Assuming $T_{\rm e}\simeq T_{\rm p} \equiv T$, $\tau$ is defined by \cite{viisanen2018}
\begin{eqnarray}
 \tau (T) &=& C_{\rm e}(T)
 \left( \frac{d\Gamma(T)}{dT} \right)^{-1}. 
\end{eqnarray}
At low $T$, from Eq.~(\ref{eq:gamma5}), one obtains
\begin{eqnarray}
 \tau_{\rm low} &=& 
 \frac{\gamma}{5\Sigma_{\rm low}N_{\rm c}\Omega_{\rm cell}T^3}
 \propto T^{-3}.
  \label{eq:tau_low} 
\end{eqnarray}
This means that when $T$ is decreased, a very small amount of energy will be exchanged between electrons and phonons due to a limited excitation of phonons. 



{\it Computational details.---}We use DFT and density-functional perturbation theory implemented into Quantum ESPRESSO code \cite{qe} to obtain $N_{\rm F}$ and $\alpha^2 F(\omega)$ in Eq.~(\ref{eq:rate_gamma}). The effects of exchange and correlation are treated within PBE-GGA \cite{pbe}. The core electrons are treated within the ultrasoft pseudopotential method \cite{uspp}. For Cu and Ag bulk calculations, the cutoff energies for the wavefunction $E_{\rm wf}$ and the charge density $E_{\rm cd}$ were set to 60 Ry and 600 Ry. For the calculation of $\alpha^2 F(\omega)$ in Eq.~(\ref{eq:coupling_e-ph}), the dense $k$-point grid of 40$\times$40$\times$40 (including $k$ and $k+q$ points), the coarse $k$-point grid of 20$\times$20$\times$20 (for constructing the charge density and the dynamical matrix), and the $q$-point grid of 10$\times$10$\times$10 are used. In case of surface calculations, $E_{\rm wf}=90$ Ry and $E_{\rm cd}=900$ Ry were used. For the calculation of $\alpha^2 F(\omega)$, the dense and coarse $k$-point grids of 24$\times$24$\times$1 and 12$\times$12$\times$1, respectively, and the $q$-point grid of 6$\times$6$\times$1 are used, which are enough to study the phonon energy range of interest. The Marzari-Vanderbilt smearing \cite{mv} with a parameter of $\sigma=0.025$ Ry is used for all calculations. 

The lattice constant is optimized to be $a_{\rm lat}=3.636$ \AA \ and 4.154 \AA \ for Cu and Ag bulk, respectively, where the total energy and forces are converged within $10^{-5}$ Ry and $10^{-4}$ a.u. For the surface calculations, where five or seven layers are considered, a vacuum layer between the surface is taken to be larger than $15$ \AA. The distance between the layers near the surface is shrinked by a few percent after a geometry optimization. The optimized volume $\Omega_{\rm cell}$, the e-ph coupling constants, $\lambda$, $\lambda\langle \omega^2 \rangle$, and $N_{\rm F}$ are listed in Table \ref{table1}. The values of $\lambda$ agree with other calculations \cite{savrasov,giri}. The magnitude of $N_{\rm F}$ is almost proportional to the number of atoms in the unit cell. 


\begin{figure}[ttt]
\center
\includegraphics[scale=0.4]{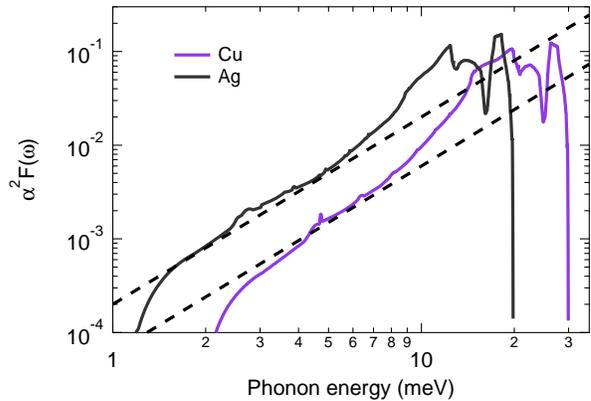}
\caption{\label{fig1} The $\alpha^2 F (\omega)$ of Cu and Ag bulk. The dashed lines indicate the curve of $\alpha^2 F (\omega) \propto \omega^2$. }
\end{figure}

{\it Bulk.---}Figure \ref{fig1} shows $\alpha^2F(\omega)$ for Cu and Ag. Below the phonon energy $\hbar\omega \simeq 8$ meV for Cu and $\hbar\omega \simeq 6$ meV for Ag, $\alpha^2F(\omega)$ shows a $\omega^2$ behavior. The deviation from a $\omega^2$ law at relatively low $\omega$ is attributed to a limited number of Brillouin zone sampling, i.e., $q$ mesh. Below, we thus use an analytical expression for low $\omega$ 
\begin{eqnarray}
\alpha^2F(\omega) = G \left( \frac{\hbar\omega}{E_0}\right)^2,
\label{eq:a2FG}
\end{eqnarray}
where $E_0=1$ meV and $\hbar\omega$ is the phonon energy in units of meV. Assuming Eq.~(\ref{eq:a2FG}), we calculate $\Gamma (T)$ in Eq.~(\ref{eq:rate_gamma}), $\Sigma_{\rm low}$ in Eq.~(\ref{eq:gamma5}), and $\tau_{\rm low}$ in Eq.~(\ref{eq:tau_low}). Table \ref{table1} lists the calculated $G$, $\Sigma_{\rm low}$, and $\tau_{\rm low}$ at $T=0.1$ K. The value of $G$ for Ag is about three times larger than that for Cu. Accordingly, $\tau_{\rm low}$ of Ag is three times shorter than that of Cu. It should be noted that $\Sigma_{\rm low}$ of Ag is larger than that of Cu by a factor of two only, due to the difference of $\Omega_{\rm cell}$ listed in Table \ref{table1}. 

\begin{figure}[ttt]
\center
\includegraphics[scale=0.4]{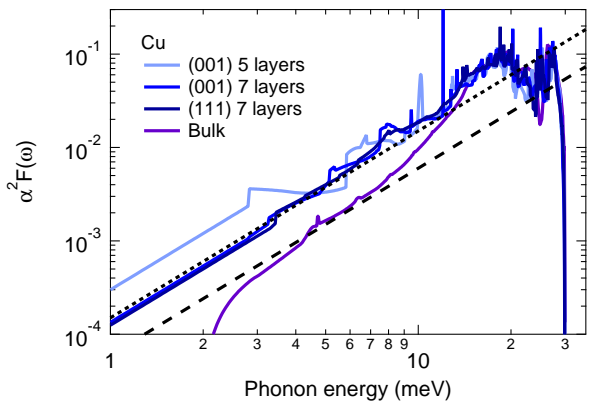}
\caption{\label{fig2} The $\alpha^2 F (\omega)$ of Cu thin films for (001) surface with five and seven layers and for (111) surface with seven layers. }
\end{figure}

\begin{figure}[ttt]
\center
\includegraphics[scale=0.4]{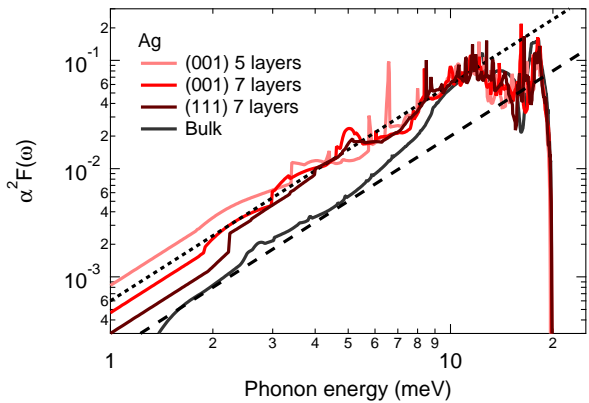}
\caption{\label{fig3} The same as Fig.~\ref{fig2} but for Ag. }
\end{figure}
\begin{figure}[ttt]
\center
\includegraphics[scale=0.4]{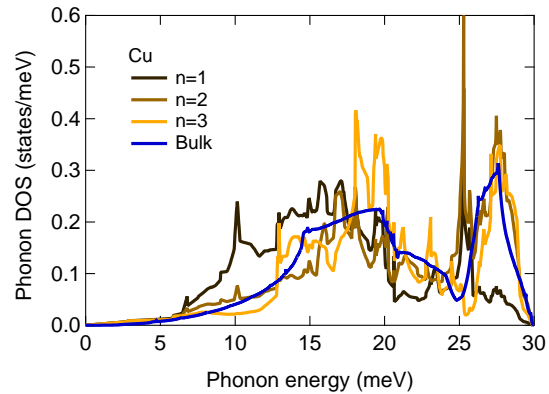}
\caption{\label{fig4} The phonon DOS for Cu bulk and partial DOS for thin film with five layers. $n$ denotes the layer number from the top ($n=1$) to the middle ($n=3$). }
\end{figure}

\begin{figure}[ttt]
\center
\includegraphics[scale=0.4]{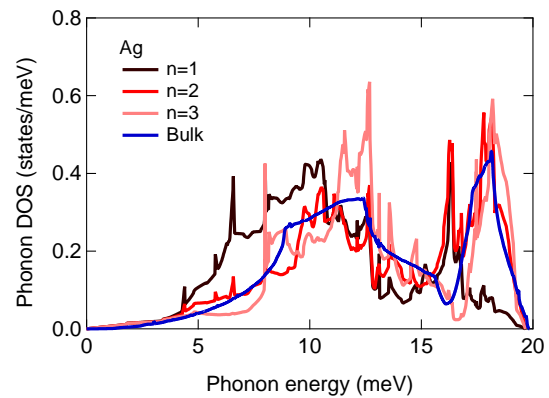}
\caption{\label{fig5} Similar to Fig.~\ref{fig4} but for Ag. }
\end{figure}

{\it Surface.---}Figures \ref{fig2} and \ref{fig3} show $\alpha^2F(\omega)$ of Cu and Ag surfaces, respectively, for the cases of (001) surface with five and seven layers and (111) surface with seven layers. It is clear that the magnitude of $\alpha^2F(\omega)$ is enhanced at low $\omega$, compared to the bulk case. The values of $G$, $\Sigma_{\rm low}$, and $\tau_{\rm low}$ at $T=0.1$ K are also listed in Table \ref{table1}. The $G$ in the surface are about three times larger than that in bulk. In accord with this enhancement, $\tau_{\rm low}$ at $T=0.1$ K becomes shorter than the bulk case: $\tau_{\rm low}\simeq 30$ $\mu$s and 7 $\mu$s for Cu and Ag surfaces, respectively, almost independent of the film thickness and crystal surface. On the other hand, the $\Sigma_{\rm low}$ are almost the same as that in bulk. This is because the enhanced $G$ and $N_{\rm F}$ are cancelled by the increased volume $\Omega_{\rm cell}$ (see Eq.~(\ref{eq:gamma5})). It does not show that $\Sigma_{\rm low}$ is correlated with $\tau_{\rm low}$. 

The enhanced $\alpha^2F(\omega)$ at low $\omega$ can be attributed to the appearance of the surface phonon mode, known as the Rayleigh mode, below the transverse acoustic phonon branch \cite{landau}. Figures \ref{fig4} and \ref{fig5} show the partial DOS for Cu and Ag thin films with five layers, respectively: $n$ denotes the layer number from the top ($n=1$) to the middle ($n=3$). The partial DOS for $(6-n)$th layer $(n=1,2)$ is exactly the same as that for $n$th layer due to the presence of the inversion symmetry against the middle layer. The DOS for the bulk is also shown. The DOS of the middle and the second ($n=2$) layers are similar to the bulk DOS at low $\omega$. On the other hand, the magnitude of the DOS of the top layer is strongly enhanced below $\hbar \omega \simeq 15$ meV (Cu) and 9 meV (Ag) compared to the bulk DOS, which can be attributed to the surface phonon mode. Similar tendency is observed for the calculations with seven layers at (001) and (111) surfaces. Similar enhancement of the phonon DOS at low $\omega$ has been reported in DFT calculations for TiC \cite{bagci} and NbC and TaC \cite{bagci2} thin films. It is natural to consider that this would modify the strength of $\alpha^2 F(\omega)$ at low $\omega$ and therefore change the e-ph dynamics at low $T$.

\begin{figure}[ttt]
\center
\includegraphics[scale=0.5]{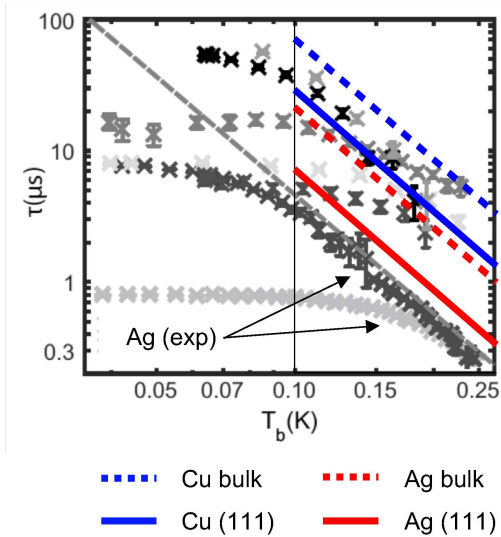}
\caption{\label{fig6} The $T$-dependence of $\tau_{\rm low}$ for Cu and Ag bulk (dashed) and (111) surface with seven layers (solid). Our results are compared to the experiment by Viisanen and Pekola, extracted from Ref.~\cite{viisanen2018}, where the plots for lower $\tau$s, indicated by arrows, are for Ag film, while the other plots are for Cu films. The different $T$-dependence of $\tau$ in Ref.~\cite{viisanen2018} is due to the use of different samples or measurements. }
\end{figure}

{\it Comparison to experiment.---}Recently, Viisanen and Pekola have investigated the e-ph relaxation dynamics at sub-Kelvin temperatures to extract the specific heat for Cu and Ag films \cite{viisanen2018}. First, they determined the value of the energy transfer rate $\Sigma_{\rm exp}$ from thermal conductance measurement: $\Sigma_{\rm exp}\simeq 2$ GW/m$^3$/K$^5$ for Cu and $\Sigma_{\rm exp}\simeq 3$ GW/m$^3$/K$^5$ for Ag. Next, they determined the relaxation time $\tau_{\rm exp}$ of $T_{\rm e}$ by investigating the response against the heating pulse: For example, $\tau_{\rm exp}\simeq 40$ $\mu$s for Cu and $\tau_{\rm exp}\simeq 3$ $\mu$s for Ag at $T=0.1$ K. Assuming a relation $\tau_{\rm exp} \propto T^{-3}$ above $T=0.1$ K, i.e., Eq.~(\ref{eq:tau_low}), they extracted the low-$T$ specific heat $\tilde{\gamma}_{\rm exp}$ (The tilde is used to denote the $\gamma$ per volume). They have found that the value of $\tilde{\gamma}_{\rm exp}$ in Ag film agrees with the free-electron estimate $\tilde{\gamma}_{\rm free}=62.4$ J/m$^3$/K$^2$, while that in Cu films is anomalously larger than the estimate $\tilde{\gamma}_{\rm free}=70.7$ J/m$^3$/K$^2$ by one order of magnitude. Below we interpret this experiment. 

Figure \ref{fig6} shows a comparison between the experimental $\tau_{\rm exp}$ \cite{viisanen2018} and the calculated $\tau_{\rm low}$ for Cu and Ag bulk and (111) surface with seven layers. The $\tau$ is significantly overestimated in the bulk calculation. With the surface effect, the value of $\tau_{\rm low}$ decreases and becomes the same order of magnitude of $\tau_{\rm exp}$. In the present calculations, no anomaly on $N_{\rm F}$ is observed except that $N_{\rm F}$ is proportional to the number of atoms in the unit cell. The estimated specific heat is $\tilde{\gamma}=100$ J/m$^3$/K$^2$ for Cu and 60 J/m$^3$/K$^2$ for Ag, agreement with $\tilde{\gamma}_{\rm free}$ above. One possible explanation for the anomalous increase in $\tilde{\gamma}_{\rm exp}$ for Cu films \cite{viisanen2018} is that the value of $\Sigma_{\rm exp}$ has been overestimated: From Eq.~(\ref{eq:tau_low}), an enhancement of $\Sigma_{\rm exp}$ would give rise to an enhanced value of $\tilde{\gamma}_{\rm exp}$ in order to fit $\tau_{\rm exp}$ correctly. In this sense, the agreement reported in Ag film \cite{viisanen2018} must be reconsidered. 

We interpret that the different $T$-dependence of $\tau$ in various Cu and Ag films in Fig.~\ref{fig6} reflects the low $\omega$ behavior in $\alpha^2 F(\omega)$. This is because, as listed in Table \ref{table1}, the magnitude of $\tau$ is inversely proportional to $G$ in Eq.~(\ref{eq:a2FG}). In addition, it is also sensitive to the randomness \cite{bergmann,echternach,sergeev,karvonen} and the boundary condition \cite{qu,hekking,cojocaru} in the sample.  

It must be noted that in most of experiments the reported values of $\Sigma_{\rm exp}$ have been a few GW/m$^3$/K$^5$ for noble metals \cite{wellstood,qu}. The value of $\Sigma_{\rm low}$ for the surface as well as bulk is smaller than that of $\Sigma_{\rm exp}$ by an order of magnitude. Furthermore, $\Sigma_{\rm low}$ is quite insensitive to the film thickness and the crystal surface (see Table \ref{table1}). It would be important to study the low $\omega$ behavior of $\alpha^2 F(\omega)$ rather than to investigate $\Sigma_{\rm exp}$ for a correct interpretation of the e-ph relaxation dynamics at low $T$. 

{\it Conclusion.---}Using DFT calculations, we have calculated the Eliashberg function $\alpha^2 F(\omega)$, the e-ph energy transfer rate $\Gamma_{\rm low}$, and the $T$-dependence of $\tau_{\rm low}$ for the bulk and surfaces of Cu and Ag. We have shown that the surface effect is strong enough to modify the magnitude of $\alpha^2 F(\omega)$ at low $\omega$, which can explain the low $T$ electron relaxation dynamics observed in a recent experiment \cite{viisanen2018}. The coupling factor usually denoted by $\Sigma$ is less correlated with $\tau$ at low $T$. We hope that the work would stimulate further study on the low $\omega$ behavior of $\alpha^2 F(\omega)$ for more realistic situations, which may further improve the discrepancy between the theory and the experiment in the e-ph relaxation dynamics. 

\begin{acknowledgments}
This study is supported by the Nikki-Saneyoshi Foundation. 
\end{acknowledgments}

\end{document}